\documentclass[preprint,aps,prc,nofootinbib]{revtex4-2}
\usepackage{color}
\usepackage{graphicx}
\usepackage{amssymb}
\usepackage{amsmath}
\usepackage{hyperref}
\usepackage[english]{babel}
\usepackage{natbib}
\usepackage[utf8]{inputenc}
\usepackage{siunitx}
\usepackage{multirow}
\usepackage{chemfig}
\usepackage{subcaption} 

\begin{document}
	 
\title{Determining the minimal mass of a proto-neutron star with chirally constrained nuclear equations of state}

\author{Selina Kunkel}
\email{kunkel@itp.uni-frankfurt.de}
\affiliation{Institut f\"ur Theoretische Physik, Goethe Universit\"at, Max-von-Laue Straße 1, D-60438 Frankfurt am Main, Germany}

\author{Stephan Wystub}
\email{deceased}
\affiliation{Institut f\"ur Theoretische Physik, Goethe Universit\"at, Max-von-Laue Straße 1, D-60438 Frankfurt am Main, Germany}

\author{J\"urgen Schaffner-Bielich}
\email{schaffner@astro.uni-frankfurt.de}
\affiliation{Institut f\"ur Theoretische Physik, Goethe Universit\"at, Max-von-Laue Straße 1, D-60438 Frankfurt am Main, Germany}

\date{\today}

\begin{abstract}
The minimal masses and radii of proto-neutron stars during different stages of their evolution are investigated.
In our work we focus on two stages, directly after the supernova shock wave moves outwards, where neutrinos are still captured in the core 
and the lepton per baryon ratio is fixed to $Y_L = 0.4$, and a few seconds afterwards,
when all neutrinos have left the star.
All nuclear equations of state used for this purpose fulfill the binding energy constraints from chiral effective field theory for neutron matter at zero temperature.
We find for the neutrino-trapped case higher minimal masses than for the case when neutrinos have left the proto-neutron star.
Thermal effects, here in the form of a given constant entropy per baryon $s$, have a smaller effect on increasing the minimal mass. 
The minimal proto-neutron star mass for the first evolutionary stage with $Y_L = 0.4$ and $s = 1$ amounts to $M_{min} \sim \SI{0.62}{M_{\odot}}$ 
and for the stage without neutrinos and $s = 2$ to  $M_{min} \sim \SI{0.22}{M_{\odot}}$ rather independent on the nuclear
equation of state used.
We also study the case related to an accretion induced collapse of a white dwarf where the initial lepton fraction is $Y_L = 0.5$ 
and observe large discrepancies in the results of the different tables of nuclear equations of state used. Our finding points towards
a thermodynamical inconsistent treatment of the nuclear liquid-gas phase transition for nuclear equations of state in tabular form
demanding a fully generalized three-dimensional Gibbs construction for a proper treatment.
Finally, we demonstrate that there is a universal relation for the increase of the proto-neutron star minimal mass 
with the lepton fraction for all nuclear equations of state used.
\end{abstract}

\maketitle
\newpage

\section{Introduction}

If the mass of a star is sufficiently high it can undergo a core-collapse supernova at the end of its evolution, from which a neutron star emerges. But before the formation of the neutron star is finished, it has to go through a hot, neutrino-opaque phase in which it is called a proto-neutron star (PNS). These objects can be described via the neutron star equation of state which has been investigated in detail, in particular in view of the recent advancements of astrophysical constraints \cite{Most:2018hfd, Koehn:2024set}. In addition to observational constraints one can infer properties of the equation of state from theoretical calculations, for example perturbative QCD at high densities \cite{Fraga:2013qra, Kurkela:2014vha}, while chiral effective field theory is a powerful tool to describe low density matter \cite{Hebeler:2013nza, Drischler:2017wtt}. 
Many equations of state have been extended to nonzero temperatures and to the case of being out of beta-equilibrium, i.e.\ for different proton fractions $Y_p$\cite{Hempel:2009mc,Hempel:2011mk,Keller:2020qhx,Keller:2022crb,Alford:2023rgp}.
As proto-neutron stars are hot, as opposed to cold neutron stars, finite-temperature equations of state are needed to determine their properties, like masses and radii. The early evolution process of neutron stars is a topic of recent interest due to observations of particularly low mass compact stars as e.g.\ the pulsar PSR J0453+1559 \cite{Martinez:2015mya} with a mass of $\SI{1.17}{M_{\odot}}$ or HESS J1731-347 \cite{Doroshenko:2022nwp} with a mass of $\SI{0.77}{M_{\odot}}$. These masses are below the lower mass limit of neutron stars emerging from modern supernova simulations of about $\sim \SI{1.2}{M_{\odot}}$ \cite{Suwa:2018uni,Stockinger:2020hse,Muller:2024aod}. 
Recently, the NICER collaboration reported the measurement of the pulsar PSR J1231-1411 with a mass of $\SI{1.04}{M_{\odot}}$
\cite{Salmi:2024bss}, which again strengthens the tension between measurements and core-collapse supernova simulations. And even though neutron star masses cannot be lower than $\SI{1.2}{M_{\odot}}$ as imposed by simulations our theoretical estimate on the lower mass limit for proto-neutron stars might still be helpful in the classification of observed compact stars with smaller masses being formed from core-collapse supernova. 

Simulations of the proto-neutron star evolution were performed as early as in 1986~\cite{Burrows:1986me}, resulting in determining the main properties of a PNS to having an approximately constant entropy per baryon over the whole star and captured neutrinos in the beginning of its evolution. The constancy of the entropy per baryon can be motivated by PNS convection \cite{1988PhR...163...51B, 1996ApJ...473L.111K}. Improved simulations from recent years, see ref.~\cite{Pascal:2022qeg, 2012PhRvL.108f1103R, 2016NCimR..39....1M, 2020MNRAS.492.5764N, 2023PhRvD.108h3040F}, have also verified the importance of PNS convection and its impact on the radial profile of the entropy and lepton fraction of the proto-neutron star. Additionally, the measurement of neutrinos correlated to the supernova SN1987A~\cite{IMB:1987klg} presented the first clear signal of a cooling PNS emitting neutrinos. Neutrino transport and convection have also been studied in refs.~\cite{Dessart:2005ck, Camelio:2017hrs, 2006A&A...457..281B} because of the dominating role the neutrino flux and diffusion plays during the proto-neutron star evolution. The presence of exotic matter like hyperons or a quark matter phase have also been considered for the equation of state, see 
refs.~\cite{Prakash:1996xs, Carter:2000xf, Weber:2004kj, Buballa:2003qv, Sagert:2008ka, Chatterjee:2015pua, Fischer:2020xjl} 
to delineate the impact of phase transitions on the evolution of a proto-neutron star.

In this work we examine minimal masses and corresponding radii of proto-neutron stars in the line of previous work \cite{Gondek:1997fd,Dexheimer:2008ax,Burgio:2010ek,Koliogiannis:2020nhh} which used different nuclear models based on Brueckner theory or a chiral SU(3) model. They found minimal masses which range from $\SI{0.43}{M_{\odot}}$ to $\SI{0.61}{M_{\odot}}$ for a PNS with an entropy per baryon of $s = 1$ and $\SI{0.56}{M_{\odot}}$ to $\SI{1.07}{M_{\odot}}$ for $s = 2$ depending on the model used for the computation. We note that some of these model calculations are not compatible with chiral effective field theory of pure neutron matter. We extend previous work by using equations of state that are compatible with the modern binding energy constraints of neutron matter from chiral effective field theory (EFT) \cite{Kruger:2013kua}, as chiral EFT gives a controlled description of low density neutron matter of up to and slightly above the saturation density of nuclear matter. It was shown in previous work that the central density  of the minimal mass configuration for the neutrino-trapped PNS stage happens to be around saturation density \cite{Gondek:1997fd}. Hence, the description with an equation of state that fulfills chiral EFT is sufficient to describe a proto-neutron stars' minimal mass. The equations of state fulfilling chiral EFT and used for this work are DD2 \cite{Hempel:2009mc, Typel:2009sy}, SFHo \cite{Steiner:2012rk} and QMC-RMF1-4 \cite{Alford:2023rgp}. 
Indeed, we find that these carefully chosen nuclear equations of state result in very similar properties of the proto-neutron star's minimal mass and corresponding radii.

We will investigate the first stage of proto-neutron star evolution by setting a constant entropy per baryon of $s = 1$ (in units of Boltzmann's constant) and for trapped neutrinos in the star with a fixed lepton fraction of $Y_L = 0.4$. Likewise we will consider the end of deleptonization, where the neutrinos have left the star and the star has a constant entropy per baryon of $s = 2$. These two stages are standard characteristic ones of a proto-neutron star evolution, see refs.~\cite{Prakash:1996xs, Janka:2017vlw},
and reflect roughly the results of more sophisticated treatments as done in \cite{Pan:2015sga, Pascal:2022qeg}. It is important to set up a baseline for the standard evolution of proto-neutron stars rooted on nuclear physics to allow for the possibility to constrain deviations of proto-neutron star evolution due to e.g.\ the appearance of new phases in the core of proto-neutron stars \cite{Prakash:1996xs, Carter:2000xf, Weber:2004kj, Buballa:2003qv, Sagert:2008ka, Chatterjee:2015pua, Fischer:2020xjl} 
as well as setting constraints on the emission of particles beyond the standard model \cite{Fischer:2024ivh, 2024arXiv240113728C}. We note that we use a constant entropy per baryon throughout the star which is born out of the observation in proto-neutron star simulations including convection,
that the proto-neutron star exhibits an isentropic profile, see the discussion above.

The outline of the paper is as follows: We will first present the theoretical framework including the equations of state adopted in this work, the implementation of the neutrino contribution, the PNS evolutionary stages and the thermal equilibrium condition. Then we present our results, where we discuss the mass-radius curves and how they change depending on the lepton fraction and the amount of entropy per baryon. 
The central densities of the minimal masses in those cases will be determined verifying that they indeed occur around saturation density. We will then extend the calculation to arbitrary lepton fractions up to $Y_L = 0.5$ where we find that the minimum proto-neutron star mass shows a clear and universal dependence on the lepton fraction $Y_L$. Moreover, we check if local thermal equilibrium is fulfilled by using the \textit{Tolman-Ehrenfest-Klein law}, which will be introduced in section \ref{TEK}.

\section{Theoretical Framework}

\subsection{Equation of State}

The nuclear equation of state (EOS) describes the thermodynamic properties of the neutron star matter under consideration. 
For this work we choose several tabulated equations of state: DD2 \cite{Hempel:2009mc, Typel:2009sy}, SFHo \cite{Steiner:2012rk} and QMC-RMF1-4 \cite{Alford:2023rgp}. There are four versions of QMC-RMF, which differ slightly by their coupling parameters. All of those equations of state have two things in common: they are relativistic mean field (RMF) models and fulfill chiral binding energy constraints at zero temperature for pure neutron matter \cite{Kruger:2013kua,Fischer:2013eka,Alford:2023rgp}.

The relativistic mean-field theory is based on the assumption that the strong interaction between two nucleons is mediated through the exchange of mesons. The Lagrangian is written down in terms of the meson fields and the baryon fields with a Yukawa coupling. 
In the mean-field approximation the mean values of the meson fields are taken, 
which involves the scalar $\sigma-$, vector $\omega$ and isovector-vector $\rho$ meson fields. The parameters are either fitted to properties of nuclei (DD2) or to nuclear matter (SFHo, QMC-RMF1-4).

Chiral effective field theory is an effective field theory that is written down in terms of chiral symmetry conserving and chiral symmetry breaking operators. 
The theoretical ansatz is an expansion in the typical momentum range of the interaction not much larger than the pion mass, where the pion is the Goldstone boson of the chiral symmetry breaking. Chiral EFT uses a systematic order-by-order approach which makes it possible to calculate two- and three-body interactions more precisely and quantify uncertainties. As the RMF approach does its expansion in lowest order approximation in relativistic many body theory it does not incorporate the power counting scheme of chiral EFT. Therefore, chiral EFT has shown to give a more robust description of neutron matter in neutron stars, where two- and three-body interactions are the ones determining the equation of state \cite{Hebeler:2013nza, Kruger:2013kua, Drischler:2017wtt}. 

Furthermore the parameter set DD2 is based on a density dependent model which is adopted from Brückner-Hartree-Fock calculations. Together with SFHo and QMC-RMF the used tables of the equations of state also incorporate nuclear statistical equilibrium
being extended to nonvanishing temperatures, which is the equilibrium between nuclei dissociating to nucleons at high temperatures and nuclei being produced at high densities. The inclusion of excluded volume effects for nuclei 
improves the description and results in a slight mutual repulsive contribution so that heavy nuclei are not overpopulated \cite{Hempel:2009mc}. 
For further information on how the equations of state were constructed and their physical background we refer to \cite{Hempel:2009mc,Typel:2009sy,Steiner:2012rk,Alford:2023rgp}. All equations of state can be downloaded at the \href{https://compose.obspm.fr/home}{CompOSE} website.

\subsection{Evolution of proto-neutron stars}

A neutron star, preceded with a proto-neutron star, marks the last stage of stellar evolution for massive stars \cite{Bethe:1990mw, 1966CaJPh..44.2553A, Colgate:1966ax}. To produce a neutron star the zero-age-main-sequence (ZAMS) mass has to be in the range of about \SIrange{8}{25}{M_{\odot}}. Only then will the mass be high enough to produce a degenerate (iron) core, where
for \SIrange{8}{10}{M_{\odot}} a degenerate \chemfig{ONeMg}-core is formed, see \cite{Kitaura:2005bt, Janka:2017vlw}.
The thermal pressure and the degeneracy pressure of the electrons are not enough to stabilize the degenerate core of the star against gravity. The photo dissociation takes away energy from the system and inverse beta decay reduces the electron pressure, so that the degenerate core collapses in free fall. The collapse is stopped by the repulsive force between the nucleons when the matter reaches a density slightly above nuclear saturation density and the infalling matter bounces back at the core forming an outgoing shock wave. 
What is left behind the shock front is a hot proto-neutron star, which over time cools by emitting neutrinos 
and becomes a cold, catalyzed neutron star.  

 As discussed in \cite{Prakash:1996xs, Janka:2012wk} neutrinos are captured in the core after core bounce, resulting in a lepton fraction somewhat less than the initial value of the degenerate core $Y_L = n_L/n_B = 0.4$, where $n_i$ is the \emph{net} number density of leptons and baryons, respectively. In this first stage 
of the proto-neutron star evolution the core is characterized by constant entropy per baryon of $s = S/A = 1$ and a mantle with a higher entropy of about $s = 5 - 10$. After about 0.5~s neutrinos start to leave the proto-neutron star and the deleptonization stage is taking place. The core heats up and arrives at an entropy of $s = 2$. At this point neutrino diffusion also comes to an end resulting in a neutrino-free proto-neutron star with a neutrino fraction of $Y_{\nu} = 0$ (note that the lepton number is not vanishing as electrons are still present in the cold neutron star). 

Another scenario of forming a proto-neutron star would be through an accretion induced collapse (AIC) of a white dwarf \cite{Fryer:1998jb, Wang:2020pzc, LongoMicchi:2023khv}. The AIC happens when the electron capture rate is high enough to reduce the central temperature and pressure, triggering the collapse. Due to a white dwarf composed of $Z/A = 0.5$ elements as helium,
carbon-12, oxygen-16, being electrically neutral the lepton fraction of a white dwarf and thus at the beginning of the PNS evolution is slightly higher than for an iron-collapse. For this reason we also examine lepton fractions up to $Y_L = 0.5$ in our parametric study to follow.

One can divide the evolution of the hot proto-neutron star into three main stages following ref.~\cite{Prakash:1996xs}: The first being shortly after core bounce at around $t = \SI{0.1}{s}$ where there is a constant lepton fraction of $Y_L = 0.4$ and constant entropy per baryon of $s = 1$. The second phase has its onset at about $\sim \SI{0.5}{s}$ with arbitrary lepton fractions $Y_L$ between $0$ and $0.4$ and arbitrary entropy per baryon $s$ between $1$ and $3$. The third stage at $\sim \SI{10}{s}$ is marked by the absence of the neutrinos, $Y_{\nu} = 0$, and a constant entropy per baryon of $s = 2$. After that thermal diffusion of neutrinos begins to take place, cooling the star's core down at a timescale of about a minute. The mantle needs approximately 100 years to cool and after doing so, the neutron star has reached its final thermalized stage and cools further by thermal neutrino emission on a timescale of about a million years. Below we will focus on the first and the third stage
of the proto-neutron star evolution.

\subsection{Ultrarelativistic Fermi gas and trapped neutrinos}

In the equations of state we use, only neutrons, protons and electrons are included. For the first PNS phase in which neutrinos are captured, we need their additional contribution to pressure, energy density, number density and entropy per baryon. Because neutrinos are almost massless we can treat them as an ultra-relativistic Fermi gas and derive their thermodynamic quantities using the standard expressions of their contribution to the equation of state as a function of the independent quantities (temperature, chemical potential, electron fraction). The pressure, net number density and entropy density are given by
\begin{eqnarray}
p &=& \frac{g}{3}T^4 \left[\frac{7\pi^2}{120} + \frac{1}{4}\left(\frac{\mu}{T}\right)^2 + \frac{1}{8\pi^2}\left(\frac{\mu}{T}\right)^4\right]\,,
\\
n &=& \frac{g}{6}T^3\left[\frac{\mu}{T} + \frac{1}{\pi^2}\left(\frac{\mu}{T}\right)^3\right]\,,
\\
\frac{S}{V} &=& gT^3 \left[\frac{7\pi^2}{90} + \frac{1}{6}\left(\frac{\mu}{T}\right)^2\right]\,,
\end{eqnarray}
where $T$ is temperature, $\mu$ the chemical potential (in this case the lepton chemical potential) and $g$ the degeneracy factor, which is $g = 1$ for (left-handed) neutrinos. For the entropy per baryon we just need to divide the entropy density $S/V$ 
by the baryon number density $n_B$ and the lepton fraction can be calculated via
\begin{equation}
Y_L = \frac{n_L}{n_B} = \frac{(n_{e^-} - n_{e^+}) + (n_{\nu_e} - n_{\overline{\nu}_e})}{(n_n - n_{\overline{n}}) + (n_p - n_{\overline{p}})} = Y_e + Y_{\nu_e}\,,
\end{equation}
where $n_L$ is the net lepton number density, including electrons, electron neutrinos and their antiparticles. We ignore contributions from muons and tauons and set the lepton numbers $Y_{\mu}$ and $Y_{\tau}$ to zero as their contribution is negligibly small because their masses are larger than the typical temperatures encountered inside a proto-neutron star. Muons could appear at relatively low densities, which nevertheless does not have a considerable impact on the properties of the star.  Similarly, the contributions of antinucleons can be ignored, too.

\subsection{Tolman-Ehrenfest-Klein law}
\label{TEK}

In 1930, Tolman \cite{Tolman:1930zza} and Ehrenfest \cite{Tolman:1930ona} investigated the role of heat within the framework of general relativity, which as an energy-equivalent should be a source of gravitation according to the equivalence principle. By assuming a static, spherically symmetric spacetime, described by the line element
\begin{equation}
    \mathrm{d}s^2 = g_{\mu \nu} \mathrm{d}x^\mu \mathrm{d}x^\nu = g_{00} \mathrm{d}t^2 + g_{ij} \mathrm{d}x^i \mathrm{d}x^j\,,
\end{equation}
and where each of the functions $g_{\mu \nu}$ depend on the spatial coordinates $x^i$ only, i.e.\ are independent of the timelike coordinate $x^0 = t$, as well as assuming a gravitating mass described by a perfect fluid in thermal equilibrium, they derived a relation not present in classical thermodynamics and/or Newtonian gravity, often referred to as the Tolman-Ehrenfest (TE) law:
\begin{equation}
    \partial_i \ln{T} = - \partial_i \ln{\sqrt{g_{00}}}
\end{equation}
or equivalently
\begin{equation}
    T\sqrt{g_{00}} = \text{const.}\,,
\end{equation}
where $T$ is the proper temperature measured by a local observer in the local rest frame and $\partial_i = \partial / \partial x^i$. The physical interpretation of this relation is that the temperature measured by a local observer depends on the gravitational field at the position of measurement, which is in stark contrast to the findings of classical thermodynamics which posits that a system in thermal equilibrium possesses a uniform temperature (the 'zeroth' law of thermodynamics).

Almost two decades later, Klein \cite{1949RvMP...21..531K} derived a generalized expression for thermodynamic consistency in curved spacetime including a chemical potential $\mu$. He finds a similar relation for the chemical potential, aptly named the Klein law, by assuming the validity of the Tolman-Ehrenfest law:
\begin{equation}
    \partial_i \ln{\mu} = - \partial_i \ln{\sqrt{g_{00}}}
\end{equation}
or
\begin{equation}
    \mu\sqrt{g_{00}} = \text{const.}\,.
\end{equation}
These two laws are widely known in the literature, oftentimes combined into the statement $\mu/T = $ constant, and are frequently considered as independent relations. However, as shown explicitly by e.g.\ \cite{Lima:2019brf}, both relations are in fact special cases of a more general relation, the Tolman-Ehrenfest-Klein (TEK) law:
\begin{equation}
    \partial_i \ln{\left( T \sqrt{g_{00}} \right)} + \frac{\mu}{T\sigma} \partial_i \ln{\left( \mu \sqrt{g_{00}} \right)} = 0\,,
    \label{eq:tek_law}
\end{equation}
where $\sigma$ denotes the specific entropy (per particle). It is easy to see that the original TE law is recovered in the cases of vanishing chemical potential, when the Klein law is conditionally fulfilled or if the ratio of chemical potential to temperature $\mu/T$ is constant.  In order to better understand the severeness of a potential violation of the law, the temperature and chemical potential are divided and thus normalized by the nuclear scale of 1 GeV. A more fundamental form of this equilibrium condition has already been used in the study of rotating hot stars, see \cite{Marques:2017zju, Goussard:1996dp} and references therein.

\section{Results}

In the following we concentrate on the first ($Y_L = 0.4$ and $s = 1$) and third ($Y_{\nu} = 0$ and $s = 2$) phase of proto-neutron star evolution. We treat them as quasi-stationary states assuming a local thermal equilibrium, where the kinetic interactions and chemical reactions are each in balance, 
i.e.\ a local kinetic and chemical equilibrium globally connected by the general TEK law.

We calculate the minimal masses and radii for those cases using the Tolman-Oppenheimer-Volkoff equations, which are first-order, nonlinear differential equations describing the radial pressure and mass.

\begin{figure}
\centering
\includegraphics[width=0.75\textwidth]{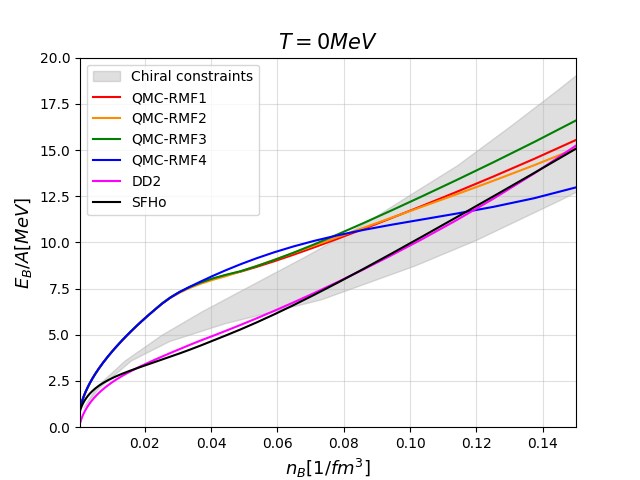}
\caption{Binding energy per nucleon against number density for $T = \SI{0}{MeV}$ in pure neutron matter for the different EOS used in this work. The gray shaded band is the region, where the binding energy per nucleon constraints of chiral EFT from \cite{Kruger:2013kua} are fulfilled.}
\label{chiralconstraints_energy}
\end{figure}

\begin{figure}
	\begin{minipage}{0.48\textwidth}
 \includegraphics[width=\textwidth]{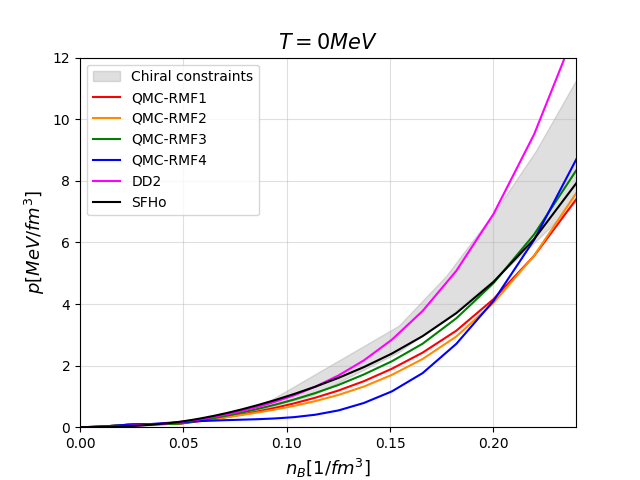}
	\end{minipage}
	\begin{minipage}{0.48\textwidth}
 \includegraphics[width=\textwidth]{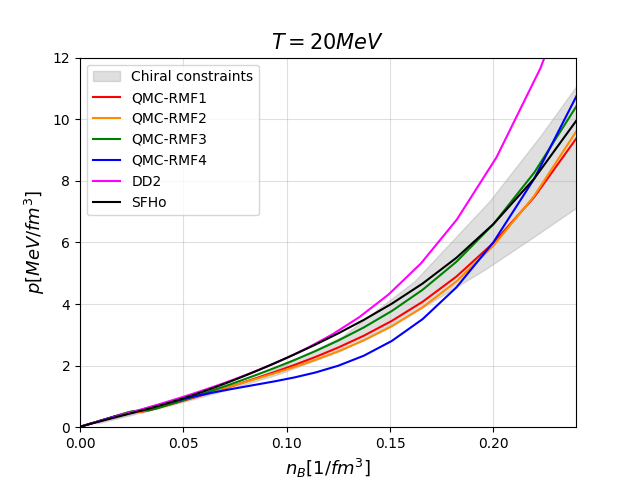}
	\end{minipage}
 \caption{Pressure against number density for $T = \SI{0}{MeV}$ (left) and $T = \SI{20}{MeV}$ (right) in neutron matter for the different EOS used in this work. The gray shaded band demarcate the pressure constraints of chiral EFT from \cite{Keller:2022crb}.}
 \label{chiralconstraints_pressure}
\end{figure}

In Figs.~\ref{chiralconstraints_energy} and \ref{chiralconstraints_pressure} we check to which extent the equations of state used in this work fulfill chiral EFT constraints for the binding energy per nucleon and the pressure of pure neutron matter taken from \cite{Kruger:2013kua} and \cite{Keller:2022crb}.

In Fig.~\ref{chiralconstraints_energy} the constraints for the binding energy per nucleon are shown in form of a gray band with the equations of state as colored lines. One observes that for densities below $\SI{0.08}{fm^{-3}}$ none of the equations of state fulfill the constraints, as opposed to the density region from $\SI{0.08}{fm^{-3}}$ to $\SI{0.15}{fm^{-3}}$.  
At and below a density of $\SI{0.08}{fm^{-3}}$, which is about half the saturation density, the transition from the core to the crust of a neutron star starts. The crust is a mixture of a lattice of nuclei immersed in a background of electrons and neutron matter, so the only relevant region of the chiral EFT constraint band for neutron stars 
is between $\SI{0.5}{n_0}$ and $\SI{1}{n_0}$, in which all of our chosen equations of state are indeed located.

Although the binding energy is the fundamental thermodynamic potential from which all quantities are derived, it is instructive to look at the recent pressure constraints from chiral EFT \cite{Keller:2022crb} in 
Fig.~\ref{chiralconstraints_pressure}, as the pressure is also the key input for the TOV-equations. In the left plot the constraints for zero temperature are shown. Only the equation of states for DD2 and SFHo are inside the chiral band for the pressure, while the EOS from QMC-RMF1-4 are too soft, especially QMC-RMF4, having too low a pressure for a given density.

In the right plot the pressure constraints for a temperature of $\SI{20}{MeV}$ are shown. In this case the equation of state for DD2 is too stiff for the relevant density region of the chiral constraint band, while SFHo and QMC-RMF1-3 fulfill the constraints for this temperature. QMC-RMF4 is still outside the chiral bounds. There seems to be a strong dependence of the equation of state on the temperature and it makes a substantial difference on which parameter one compares with in regards to how much the equations of state comply with the constraints from chiral EFT. Nevertheless, all of the depicted equations of state are in a reasonable range around the chiral band to be considered in the following for delineating the uncertainty of the underlying nuclear equation of state for the calculation of proto-neutron star properties. 

\begin{figure}[t]
\hspace*{-0cm}
\centering
\includegraphics[width=0.85\textwidth]{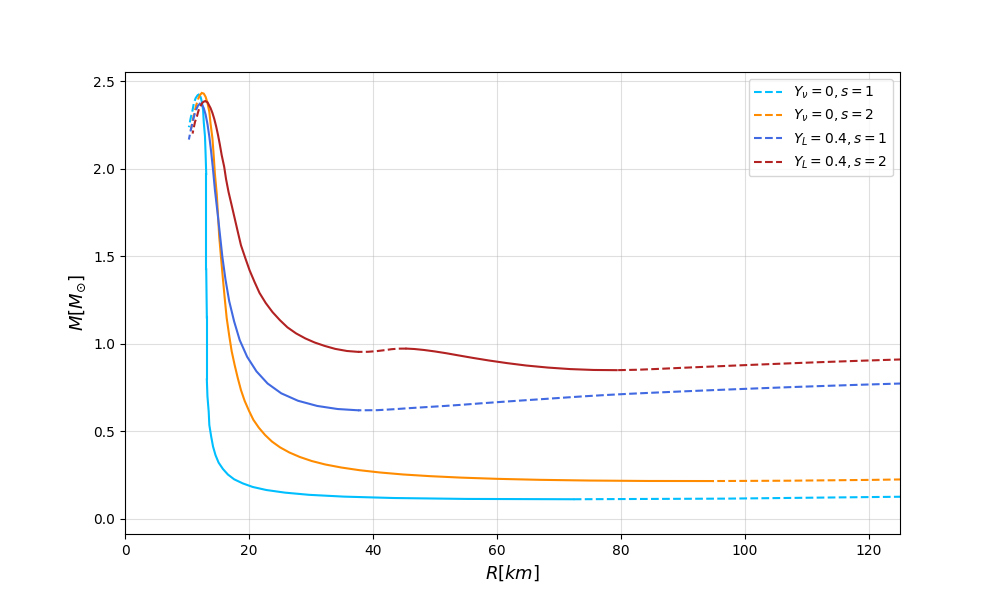}
\caption{Mass-radius curves for a neutrino-free and neutrino-trapped proto-neutron star, each for a constant entropy per baryon of $s = 1$ and $s = 2$ for DD2. The unstable mass-radius configurations are marked by dashed lines.}
\label{mrcomparison}
\end{figure}

By using the TOV-equations and the above mentioned equations of state we can calculate the mass-radius curves and determine the minimal masses. Unlike cold, catalyzed neutron stars, proto-neutron stars do not have a vanishing pressure at the surface, but a pressure that has a nonzero value as there is still a thermal halo around the star during evolution. For the surface pressure we choose as the cutoff $p = \SI{e-6}{MeV/fm^3}$ as done in ref.~\cite{Pascal:2022qeg} and checked that our results are insensitive to the detailed choice of this pressure. Fig.~\ref{mrcomparison} shows the mass-radius curves for different lepton fractions and different constant ratios of entropy per baryon using DD2. Note, that the brown and light blue lines are for choices of the lepton fraction $Y_L$ and the entropy per baryon $s$ outside the ranges usually considered for proto-neutron stars to delineate the impact of thermal effects and the abundance of neutrinos on the mass-radius curves on a wider parameter scale.

It is immediately apparent that the amount of neutrinos is the key factor for a higher minimal mass. The minimum mass $M_{min}$ configuration has a smaller radius for a high lepton fraction of $Y_L=0.4$ ($\sim R_{min} = \SI{38}{km}$) 
compared to the neutrino-free case ($\sim R_{min} = \SI{80}{km}-\SI{100}{km}$). The reason for the smaller radius in the neutrino-trapped case is the additional neutrino pressure. Although the neutrino pressure is small on nuclear scales, it makes a difference at the low central pressures of the minimal mass configurations. For the maximum mass on the other hand the presence of neutrinos result in a lower maximum mass configuration. With increasing lepton fractions the number of electrons rises as well. In consequence, to maintain charge neutrality, more protons have to be present. This increase in proton number makes the star more isospin symmetric and lowers the energy of the nucleons. As a consequence, the pressure is lowered and the maximum supportable mass shrinks \cite{Dexheimer:2008ax}. Hence, the addition of neutrinos softens the equation of state. 
We note in passing that the opposite behavior has been observed when one adds exotic matter, like hyperons or quark matter \cite{Keil:1995hw,Prakash:1995uw,Glendenning:1994za}. 
With more neutrinos, i.e.\ less neutrons, the formation of exotic matter is less favored. The softening due to the additional degrees of freedom or a new phase is not present at this stage and can only appear later in the evolution of the proto-neutron star during deleptonization. The delayed softening of the EOS can result in a delayed collapse to a black hole, as the maximum mass decreases when exotic matter appears in the core of the deleptonized and cold neutron star \cite{Baumgarte:1996iu}. 

Furthermore thermal effects also play a role making proto-neutron stars in general more massive, although for high central densities this effect is less pronounced compared to low central densities. Similarly, the impact of the presence of neutrinos on the maximum masses, i.e.\ the difference for the maximum mass between $Y_L = 0.4$ and $Y_{\nu} = 0$, is modest.

\begin{figure}
\hspace*{-0cm}
\centering
\includegraphics[width=0.85\textwidth]{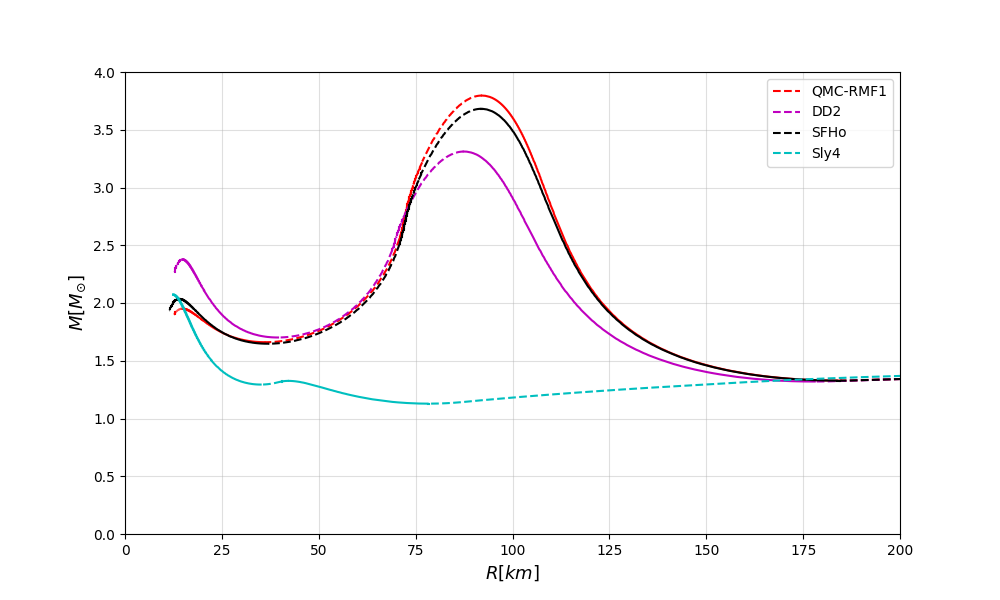}
\caption{Mass-radius curves for equations of state with a constant entropy of $s = 2$ and a lepton fraction of $Y_L = 0.5$. 
The unstable parts are marked by dashed lines. There appears a large second maximum in the mass-radius relation at large radii which is unphysical due to the interpolation  at the liquid-gas phase transition.}
\label{mr_yl05}
\end{figure}

The stability of configurations in the mass-radius diagram is determined by the eigenfrequencies of the normal radial modes, which can be calculated via an iterative integration of a Sturm-Liouville eigenvalue equation, see \cite{1966ApJ...145..505B}. It has been proven in this work that one can simply look at the dependence of the mass and the radius on the central energy density for judging the stability of the configuration. For stable configurations the mass should always increase with increasing energy density while the radius should decrease. 
So every time the mass-radius curve turns counter-clock-wise with increasing central energy density, the configurations switch from a stable to an unstable branch. Clock-wise turns in the mass-radius curve 
indicate the transition from an unstable branch to a stable one. The unstable region after the former condition is fulfilled (i.e.\ at the maximum mass) or the region between the former and the latter turning points are marked by dashed lines in Fig.~\ref{mrcomparison}. Although these stability conditions are derived for the case of vanishing temperatures, they apply to a rather good approximation also to the isothermal or isentrope neutrino-trapped cases, see ref.~\cite{Gondek:1997fd}.

One notices that for high lepton fraction and high entropy a twin star configuration appears, where there exists stable configurations for the same mass but different radii \cite{Gerlach:1968zz}. Such a case is present when the
mass-radius curve exhibits an unstable region followed by a stable one up to a second maximum. 
The new stable branch is called the third family of compact stars, the second being the neutron star branch and the first the white dwarf branch. Such a feature is present in the mass-radius curve 
if the equation of state has a first order phase transition or a nearly constant pressure as a function of energy density followed by a steep increase of the pressure at higher energy density.

\begin{figure}[t]
\hspace*{-0cm}
\centering
\includegraphics[width=0.75\textwidth]{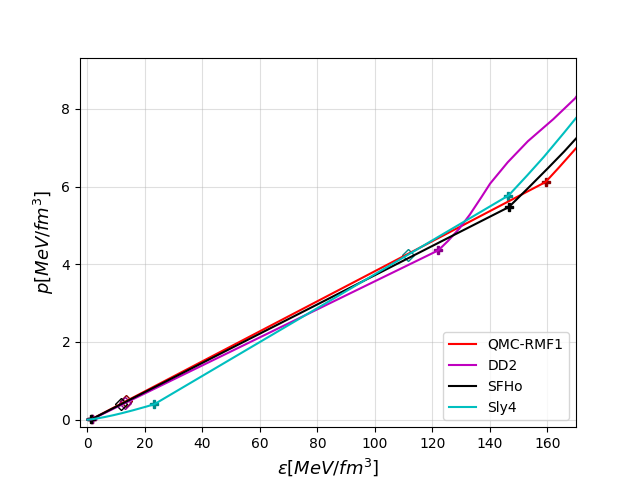}
\caption{Pressure against energy density for the case $Y_L = 0.5$ and $s = 2$ for different equations of state. The diamonds indicate the central pressures and densities of the maximal masses of the additional branches. The crosses mark the beginning and end of the transition region.}
\label{p(e)}
\end{figure}

The third stable branch has been studied for a transition to quark matter \cite{Kampfer:1981yr, 2000A&A...353L...9G, Schertler:2000xq, Zacchi:2016tjw, Christian:2017jni} and hyperon matter \cite{Schaffner-Bielich:2000nft}. In our case the phase transition in question leading to the additional stable branch is the nuclear liquid-gas phase transition. At half the saturation density the crust, being made out of nuclei, switches to the liquid core, which consists of neutrons, protons, electrons and in our case additionally neutrinos. 
This phase transition proceeding along an isentrope is responsible for a jump in the pressure and a jump in the energy density resulting in an unstable mode. There could be a mixed phase present in this region in which bubbles form and in which convection takes place in order to maintain a constant entropy per baryon in the mixed phase. The onset of the third stable branch marks the end of this transition region. The jump in energy density is small for low lepton fractions, but increases for higher lepton fractions and higher entropy. The corresponding maximum masses of the first (neutron star) branch can be extremely high 
for high lepton fractions as one can observe in Fig.~\ref{mr_yl05} for $s = 2$ and $Y_L = 0.5$.
For $Y_L=0.5$ and an entropy per baryon of $s = 2$ most of the used equations of state generate in the additional stable branch a very high maximal mass of around $\sim \SI{3.5}{M_{\odot}}$, see Fig.~\ref{mr_yl05}. 
Note, that this high maximum mass does not violate causality as the maximum is at very large radii. 
We include the Sly4 equation of state for comparison \cite{Chabanat:1997qh} which also shows a twin star solution. 
The Skyrme equation of state Sly4 represents a unified equation of state which uses the same description for the crust and the core, making the jump in energy density smaller compared to the other equations of state which are not a unified one. In addition it is also constrained by the equation of state of pure neutron matter.
The origin of the large maximal mass for the non-unified equations of state becomes clear when one looks at the equation of state in the transition area depicted in Fig.~\ref{p(e)}. 
As the entropy is kept constant and not the temperature, the transition does not take the form of a jump in energy density at a constant pressure, but rather involves a drastic increase in the pressure. 
The high mass branch is therefore an artifact of the linear interpolation in the  mixed phase between the low-density (crust) and high-density (core) phases for a constant entropy per baryon. It can be seen that the start of the transition for Sly4 begins at a higher density and the interpolation involves a smaller range in density, which results in a smaller 
maximum mass of the first branch corresponding to a higher central pressure compared to the other non-unified
equations of state. Nevertheless, the unphysical bump of the mass-radius curve at large radii 
seems to be an universal feature of chirally constrained equations of state in tabular form, even for unified ones.

It is important to note, that the appearance and height of the twin star branch depends heavily on the kind of interpolation adopted in the region of the phase transition. A linear interpolation in $\log (p)$ and $\log(\epsilon)$ for example would result in not having an additional branch in Fig.~\ref{mrcomparison} for $s = 2$ and $Y_L = 0.4$. The minimal mass for this curve would decrease by $\sim \SI{0.13}{M_{\odot}}$ and shift to the right by about $\SI{10}{km}$. Aside from that the twin branch in Fig.~\ref{mr_yl05} for $Y_L=0.5$ is still present but the mass-radius curve is lower, the maximum of the first branch being at $\sim \SI{1.4}{M_{\odot}}$ and the minimum at $\sim \SI{1.2}{M_{\odot}}$. So for higher lepton fractions and entropies the interpolation method seems to make a difference. 
However, if one wants to treat the nuclear liquid-gas phase transition correctly, one would have to calculate the phase coexistence surface in the whole phase space of temperature $T$, baryon chemical potential $\mu_B$, and charge chemical potential $\mu_Q$ and perform a generalized Gibbs construction in three dimensions as outlined in ref.~\cite{Hempel:2009vp}.
Note, that the data tables available on CompOSE do not allow for making such a generalized Gibbs construction, 
so that one has to perform an interpolation in the mixed phase. Unfortunately, these data tables are used
as input for supernova and neutron star merger simulations implying the possibility of producing artefacts
as the liquid-gas phase transition is not handled in a thermodynamical consistent way. Intriguingly, 
this seems to have been overlooked in previous simulations but opens the tantalizing prospect of forming
macroscopic bubbles in the nuclear liquid-phase transition at low density and high lepton fractions 
providing a source of gravitational waves at the MHz frequency scale. The nuclear bubbles can be of macroscopic sizes as several chemical potentials are involved in total so that the bubbles can be charge neutral but with a higher baryon density and a higher lepton fraction compared to the background matter, as discussed in \cite{Pagliara:2009dg} for the quark-hadron phase transition.

As mentioned above the phase transition gets more pronounced for higher $s$ and higher $Y_L$. This behaviour can be understood by imagining following the isentropes in the nuclear phase diagram. 
When one reaches the transition hypersurface from low densities, one has to move a larger distance in phase space for a given higher entropy value along the hypersurface in the $T,\mu_B,\mu_Q$-diagram to reach the corresponding isentrope in the high-density phase. 
Meanwhile a higher lepton fraction results in having more protons and less neutrons in the neutron star matter. It is well known that compared to neutron matter where the binding energy per nucleon is monotonically increasing with density, symmetric matter has a minimum in the binding energy at saturation density, which leads to an instability in the equation of state. This instability increases with increasing lepton fraction, i.e.\ proton fractions, and causes larger instabilities in the mass-radius curve. Previous work on thermal twins with non-zero temperature and non-zero lepton fraction show the same behavioral pattern, even if it was always the quark-hadron phase transition and not the nuclear liquid-gas phase transition which has been studied, see e.g.\ refs.~\cite{Hempel:2015vlg, Carlomagno:2023nrc}. 

\begin{figure}
	\begin{minipage}[t]{0.49\textwidth}
 \includegraphics[width=\textwidth]{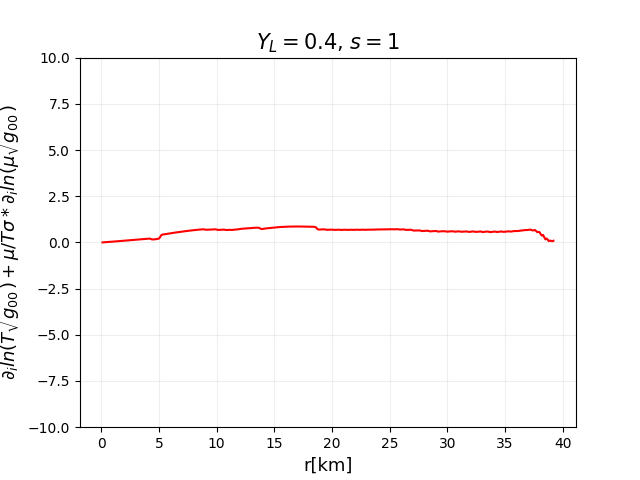}
	\end{minipage}
	\begin{minipage}[t]{0.49\textwidth}
 \includegraphics[width=\textwidth]{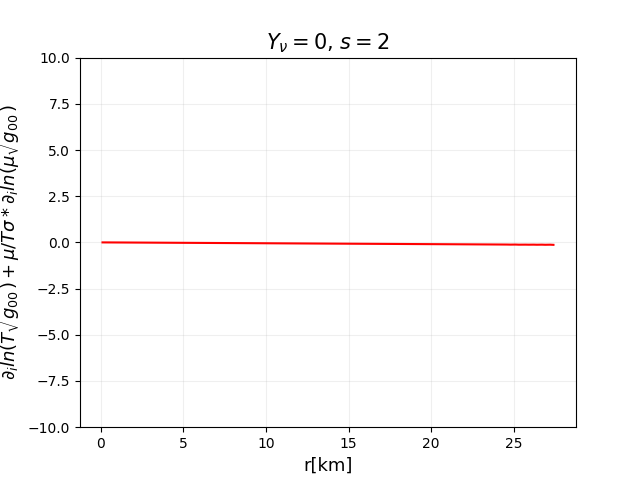}
	\end{minipage}
 \caption{Tolman-Ehrenfest-Klein law for the first (left plot) and third (right plot) proto-neutron star evolutionary stage of the minimal mass configuration using the EOS of QMC-RMF1.}
 \label{teklaw}
\end{figure}

That there is a thermodynamic inconsistency in the treatment of the liquid-gas phase transition for proto-neutron stars can be manifested by looking at the Tolman-Ehrenfest-Klein law, see Eq.~(\ref{eq:tek_law}).
In Fig.~\ref{teklaw} the results of the Tolman-Ehrenfest-Klein law using QMC-RMF1 are shown for the first and third evolutionary stages of the minimal mass proto-neutron star. While the law is fulfilled for the neutrinoless stage, being at zero value, it is severely  violated in a certain radial region for the neutrino-trapped stage. It starts being at a nonzero, positive value at around $\sim \SI{7}{km}$. Exactly at this point we are at $\SI{0.5}{n_0}$, which is the crust-core transition. As discussed earlier there is a jump in energy density for $Y_L = 0.4$, which is dealt with by interpolation. This approximation is responsible for the TEK-law not being entirely fulfilled for the minimal mass configuration of the first PNS stage. If one would treat the phase transition thermodynamically correct by performing a full three-dimensional Gibbs construction the TEK-law should be fulfilled also for this case as it is simply an expression of thermodynamic consistency.

The minimal masses, associated radii and central densities of both cases are listed in tables (\ref{firstphasetable}) and (\ref{thirdphasetable}) for all equations of state and for the two phases studied of proto-neutron star evolution. They represent the case in which the interpolation is done linearly in pressure and energy density. We also did an interpolation in $\log(p)$ and $\log(\epsilon)$ and confirm that those values only change at the percentage level.

\begin{centering}
\begin{table}
\begin{tabular}{ |p{3cm}||p{3cm}|p{3cm}|p{3cm}|  }
 \hline
 \multicolumn{4}{|c|}{$Y_L = 0.4$, $s = 1$} \\
 \hline
 EOS & $M_{min}[M_{\odot}]$ & $R[km]$ & $n_B[fm^{-3}]$\\
 \hline
 QMC-RMF1   & 0.617 & 37.9 & 0.149\\
 QMC-RMF2   & 0.615 & 36.6 & 0.163\\
 QMC-RMF3   & 0.618 & 38.0 & 0.145\\
 QMC-RMF4   & 0.624 & 36.1 & 0.164\\
 DD2        & 0.620 & 38.3 & 0.134\\
 SFHo       & 0.613 & 37.0 & 0.153\\
 \hline
 \end{tabular}
  \caption{Masses, radii and central densities of the minimal mass configuration for the first PNS phase with $Y_L = 0.4$ and $s = 1$.}
 \label{firstphasetable}
\end{table}

\begin{table}
\begin{tabular}{ |p{3cm}||p{3cm}|p{3cm}|p{3cm}|  }
 \hline
 \multicolumn{4}{|c|}{$Y_{\nu} = 0$, $s = 2$} \\
 \hline
 EOS & $M_{min}[M_{\odot}]$ & $R[km]$ & $n_B[fm^{-3}]$\\
 \hline
 QMC-RMF1   & 0.238 & 101 & 0.00821\\
 QMC-RMF2   & 0.238 & 101 & 0.00823\\
 QMC-RMF3   & 0.238 & 101 & 0.00821\\
 QMC-RMF4   & 0.238 & 101 & 0.00821\\
 DD2        & 0.216 & 94.9 & 0.0121\\
 SFHo       & 0.195 & 85.0 & 0.0161\\
 \hline
\end{tabular}
 \caption{Masses, radii and central densities of the minimal mass configuration for the third PNS phase with $Y_{\nu} = 0$ and $s = 2$.}
 \label{thirdphasetable}
\end{table}
\end{centering}

Looking at the first PNS phase (table (\ref{firstphasetable})) one can observe that the results do not significantly differ between the different equations of state used. 
All minimal masses are around $ \sim \SI{0.62}{M_{\odot}}$, radii are all around $\sim \SI{38}{km}$ and the central densities at which those minimal mass configurations occur are approximately at saturation density. This tells us that the calculated properties are fairly reliable, because we know the equation of state up to saturation density very accurately by virtue of chiral EFT calculations. Even though we see from Fig.~\ref{chiralconstraints_pressure} that the constraints are not entirely fulfilled for each equation of state at each temperature the addition of neutrinos and the consideration of a constant entropy per baryon result in all six equations of state being very similar to each other and generating nearly the same results for the minimal mass of a proto-neutron star. 

After neutrinos left and the star heats up (table (\ref{thirdphasetable}) the minimal mass configuration is lighter in comparison ($M \sim \SI{0.2}{M_{\odot}}$) but has a larger radius which depends on the chosen equation of state. The number densities are again at the same order of magnitude, but considerably 
lower than in the first phase of proto-neutron star evolution, which explains the smaller mass to radius ratio. One notices that the results for the QMC-RMF equations of state are almost identical, the reason being that they all use the HS(IUF) equation of state for the crust and thus generating identical numbers at low densities.
The differences between the calculated properties compared to the other  equations of state are also negligible for the masses but not for the radii. Here we observe a discrepancy of $\SI{15}{km}$ which could be related to the fact that the equations of state show some differences at low baryon densities. While the crust EOS at zero temperature is well known and terminates around half saturation density, the crust of a neutron star melts at nonvanishing temperature so that the property of the adopted hot and low density EOS, i.e.\ below the crust-core transition density, are becoming relevant. 

\begin{figure}
\hspace*{-0cm}
\centering
\includegraphics[width=0.75\textwidth]{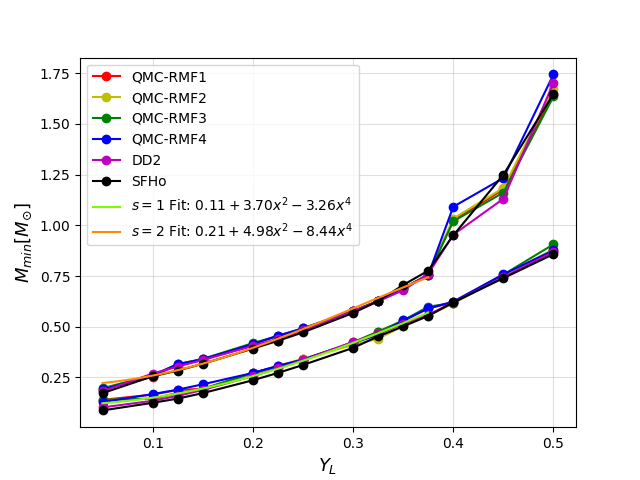}
\caption{The minimal mass dependence on the lepton fraction for different ratios of the entropy per baryon $s$. 
The lower curve corresponds to $s = 1$ and the upper curve to $s = 2$. 
The minimal mass increases uniformly with the lepton fraction $Y_L$. The interpolation method at the crust-core transition is chosen to be linear in pressure.}
\label{mmin_yl}
\end{figure}

While the minimal mass for all equations of state seems to be the same for both evolution stages we are working with, it is also of interest to take a look at the dependence on the lepton fraction as well. Fig.~\ref{mmin_yl} shows the dependence of the minimal mass of a proto-neutron star on the lepton fraction.
One notices that thermal effects increase the mass, so that the curves for $s = 2$ are generically
at higher masses compared to the case $s=1$. As expected, the minimal mass increases with higher lepton fractions
because of the higher additional neutrino pressure. 
The relation of the minimal mass with the lepton fraction is not linear as we can see for low and high lepton fractions, but follows a nonlinear relation.
The slope increases for higher lepton fractions for both case shown, $s=1$ and $s=2$.

Lowering the lepton fraction $Y_L$, the minimal mass does not approach zero, but rather goes asymptotically towards the beta-equilibrium limit. In beta-equilibrium the lepton chemical potential is set to zero, so that the net neutrino number density vanishes. This corresponds to the case of an overall small lepton fraction. At small temperatures the fixed $Y_L$ is almost only given by the electrons, which also only make up a few percent at most. So in both cases the number of electrons and leptons is very similar, so that they can be easily compared. Note that we can never reach $Y_L = 0$ because there have to be electrons around in the neutron star. The range of minimal masses at $Y_L = 0.1$ for $s = 1$ is also comparable to the lower mass limit found for cold neutron stars at zero temperature with $M_{min} \sim \SI{0.1}{M_{\odot}}$. 
This finding relates to the fact that almost no neutrinos are present and one is close to the zero temperature limit at the chosen entropy per baryon. 
As noted above the slope increases for higher lepton fractions, the effect being higher for higher entropy. These features can be correlated to the first order liquid-gas phase transition mentioned before, which gets more pronounced for higher lepton fraction and entropy. However, we note that the upper end of the curve may look different depending on the type of interpolation used for the liquid-gas phase transition at high lepton fractions. 
Hence, the results for large lepton fractions, definitely above $Y_L = 0.4$, should be taken with care. We test the changes in the $M_{min}$-$Y_L$ relation for a logarithmic interpolation using the EOS of DD2 and find visible differences for $s=1$ at $Y_L=0.5$ and for $s=2$ for $Y_L=0.4$ and above. 
The curves shifts to lower values being $M_{min}=0.82M_\odot$ for $s=1$ at $Y_L=0.5$ and 
for $s=2$ we find $M_{min}=0.83M_\odot$ at $Y_L=0.4$ and $M_{min}=1.2M_\odot$ at $Y_L=0.5$ using the logarithmic interpolation.

\begin{figure}
\hspace*{-0cm}
\centering
\includegraphics[width=0.75\textwidth]{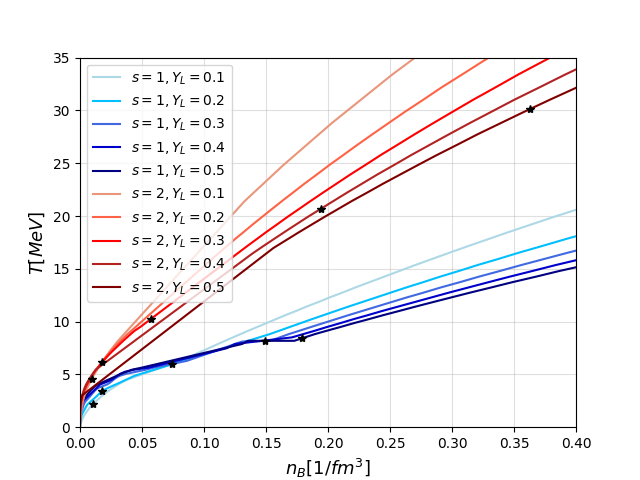}
\caption{Isentropes as functions of temperature against number density for $s = 1$ (blue curves) and $s = 2$ (red curves) for lepton fractions from $Y_L = 0.1$ to $0.5$. The used equation of state is QMC-RMF1. The central properties of the minimal mass configurations are marked by black stars.}
\label{isentropes}
\end{figure}

The curves for all equations of state seem to align almost perfectly except for large lepton fractions 
for the case $s = 2$. One reason for the discrepancy at high lepton fractions and high entropies might originate from the the big energy density jump of the liquid-gas phase transition at large values of $Y_L$, which can result in high central densities of up to $\sim \SI{2.5}{n_0}$. At these densities the differences of the 
different models of the equations of state used become more pronounced in contrast to the similarities at lower densities. These differences can be 
delineated by looking at the isentropes of temperature against number density in Fig.~\ref{isentropes}. The black stars mark the values for the minimal mass configuration.
One observes that the location of the minimal mass $M_{min}$ follows a certain pattern lying at increasing density for increasing lepton fractions.
For $s = 2$ the points of the minimal mass for $Y_L = 0.4$ and $Y_L = 0.5$ are at very high densities compared to lower lepton fractions. 
So for those cases the differences between the minimal masses for the different EOS used become pronounced. 
By using a logarithmic interpolation the minimal mass configurations (the black stars in the figure) would shift to lower densities. 
For low lepton fractions up to $Y_L = 0.3$ the effect is found to be negligible. 
The position of the corresponding ($s = 1, Y_L = 0.4$) and ($s = 1, Y_L = 0.5$) minimal mass configurations shift by about 
$\SI{0.02}{fm^{-3}}$ for the logarithmic interpolation compared to the linear one. 
For ($s = 2, Y_L = 0.4$) and ($s = 2, Y_L = 0.5$) however, the location of the black stars move to the left in the figure by $\SI{0.1}{fm^{-3}}$.

Nevertheless, one observes a universal relation for $M_{min}(Y_L)$ for proto-neutron stars, 
which seems to be rather independent on the equation of state used. The green and orange curves in Fig.~\ref{mmin_yl} show a polynomial fit to the computed values
of the form
\begin{equation}
    M(Y_L) = a + b \cdot Y_L^2 + c \cdot Y_L^4\,,
\end{equation}
with the parameters listed in table~\ref{fitparameterstable}. We stress that the fit only includes the data points 
up to a lepton fraction of $Y_L = 0.375$ as the minimal masses up to this point do not change significantly for the two interpolation methods used as discussed above.

\begin{table}
\begin{tabular}{ |p{3cm}||p{3cm}|p{3cm}|p{3cm}|  }
 \hline
 S/A & a & b & c\\
 \hline
 s = 1   & 0.11 & 3.70 & -3.26\\
 s = 2   & 0.21 & 4.98 & -8.44\\
 \hline
\end{tabular}
 \caption{Fit parameters for the equation $M_{min}(Y_L) = a + b \cdot Y_L^2 + c \cdot Y_L^4$ in units of $M_\odot$, which describes the minimal mass of a PNS as a function of $Y_L$ in Fig.~\ref{mmin_yl} up to $Y_L=0.375$}.
 \label{fitparameterstable}
\end{table}


\section{Summary and Conclusion}

In this work we investigated the mass-radius properties of proto-neutron stars at different points in their evolutionary stage using six different relativistic mean field equations of state. We studied proto-neutron stars for a non-zero constant entropy per baryon and trapped neutrinos at the start of their evolution. For parmetrically studying the properties of proto-neutron stars 
we worked with constant ratios of entropy per baryon of $s = 1$ and $s = 2$ and different lepton fractions $Y_L$. 
We started our investigation by checking if the used equations of state not just fulfill the binding energy constraints of chiral EFT at $T = \SI{0}{MeV}$, but also the pressure constraints for $T = \SI{0}{MeV}$ and $T = \SI{20}{MeV}$. Surprisingly, the EOS for DD2 and SFHo are lying within the $T = \SI{0}{MeV}$ chiral EFT band but the EOS of DD2 does not in the $T = \SI{20}{MeV}$ band. 
On the other hand the QMC-RMF models are not lying in the $T = \SI{0}{MeV}$ chiral EFT band while being in the $T = \SI{20}{MeV}$ chiral EFT band even though they were fitted to binding energy constraints at $T = \SI{0}{MeV}$ \cite{Alford:2023rgp}. 
Despite these slight deviations from the chiral EFT constraints we find that all equations of state 
show similar results in the calculations for proto-neutron stars. 

The mass radius curves for proto-neutron stars show a sizable dependence on the value used for the entropy and the lepton fraction. We find that with a lepton fraction of $Y_L = 0.4$ the minimal mass is larger compared to the case without neutrinos because the additional neutrino pressure is not negligible at these low central densities and thus the star is able to stabilize more mass against the pull of gravity. 
The effect of neutrinos for the maximal mass is the opposite. The maximal mass is reduced because of the lowered neutron pressure when adding neutrinos. Adding thermal effects, for a given entropy per baryon, increases all masses but to a lesser extent compared to adding neutrinos. For high lepton fractions and ratios of entropy per baryon twin star solutions are found due to the liquid-gas phase transition of nuclear matter. However, the appearance of this additional branch depends strongly on how one interpolates in the mixed phase region of the phase transition. All equations of state show similar results for the two characteristic phases of proto-neutron star evolution investigated. Based on our calculations we determine the minimal mass and radius for the two characteristic phases of proto-neutron star evolution.
The first phase ($Y_L = 0.4, s = 1$) has a minimal mass of $M_{min} \sim \SI{0.62}{M_{\odot}}$ and $R \sim \SI{38}{km}$. For the third phase ($Y_{\nu} = 0, s = 2$) we get $M_{min} \sim \SI{0.22}{M_{\odot}}$ and $R \sim \SI{90}{km}$. The corresponding central density for the neutrino-trapped phase is around saturation density while the one for the neutrinoless phase is considerably lower and around $n_B \sim \SI{0.01}{fm^{-3}}$. Our computed result for the minimal mass of the first phase exceeds previous calculations from \cite{Gondek:1997fd, Dexheimer:2008ax, Burgio:2010ek, Koliogiannis:2020nhh}. By employing chiral effective field theory, we provide a more precise and reliable estimate for the range in which the minimal masses of proto-neutron stars lie. However, there remains room for improvement in determining these properties, particularly in the interpolation between crust and core at low densities, which should be performed using a Gibbs construction. 

We checked for local thermal equilibrium by confronting our calculations with the Tolman-Ehrenfest-Klein law. We find that the TEK law is indeed fulfilled for the neutrinoless case. However, for the neutrino-trapped case we find that the TEK law is not entirely fulfilled because of the interpolation used for the nuclear liquid-gas phase transition. The calculation of a thermodynamically consistent full three-dimensional Gibbs construction is needed to ensure thermodynamic consistency. 
We note that this thermodynamic inconsistency only appears for high lepton fractions
when the proton fraction becomes so large that the matter enters the spinodal instability of the nuclear liquid-gas phase transition. That the liquid-gas phase transition can be present in proto-neutron star evolution opens the tantalizing perspective that macroscopic bubbles can form which can generate gravitational waves by bubble collisions.

We further studied the dependence of the minimal mass on the lepton fraction and showed that there is a universal increase of the minimal mass with the lepton fraction. The slope of the relation increases for higher lepton fractions due to an increasing jump in the energy density from the liquid-gas phase transition.
We observe a similar behavior for all equations of state used and provide a common fit formula for both entropy cases 
($s=1$ and $s=2$) studied.

In summary we showed that proto-neutron stars' properties are highly sensitive to the chosen constant entropy and lepton fraction, where the appearance of the liquid-gas phase transition is particularly affected by this choice. 
In the future, the mixed phase should not be just approximated by an interpolation, but must be modeled with a general Gibbs construction involving three chemical potentials, which in addition also includes non-zero temperatures. Such a calculation inquires the explicit calculation of the nuclear equation of state in its full three-dimensional phase space from a microscopic model.
Nevertheless, we find similar results for the equations of state studied being within or close 
to the chiral EFT results. One could make use of the error band of the chiral EFT equation of state for constant entropy and lepton fraction. With this one can calculate an error band for the minimal masses of proto-neutron stars with high precision. In addition the relation found for the minimal mass as a function of lepton fraction can be rigorously determined from first principle chiral EFT calculations.

\begin{acknowledgments}
 We thank Hosein Gholami, Veronica Dexheimer, Micaela Oertel for helpful discussions and comments. We thank in particular Thomas Janka for his critical reading of the introduction, help on providing references related to and insightful discussions on proto-neutron star evolution from supernova simulations and AICs. 
 Stephan Wystub was supported by the F\&E program of GSI Helmholtzzentrum für Schwerionenforschung Darmstadt, Germany.
 The authors acknowledge support by the Deutsche Forschungsgemeinschaft (DFG, German Research Foundation) through the CRC-TR 211 'Strong-interaction matter under extreme conditions' -- project number 315477589 -- TRR 211. 

\end{acknowledgments}

%


\end{document}